\documentclass[aps,prl,twocolumn,amsmath,amssymb]{revtex4}
\usepackage{graphicx}
\usepackage{dcolumn}
\usepackage{bm}
\usepackage[usenames]{color}

\begin{document}

\title{The Effects of Substrate Phonon Mode Scattering on Transport in Carbon Nanotubes}

\author{Vasili Perebeinos$^{1*}$, Slava Rotkin$^2$, Alexey G. Petrov$^3$, and Phaedon Avouris$^1$ \\
$^1$IBM Research Division, T. J. Watson Research Center, \\ Yorktown
Heights, New York 10598 \\
$^2$Physics Department, Lehigh University, 16 Memorial Dr. E.,
Bethlehem, PA 18015 \\Center for Advanced Materials and
Nanotechnology, \\ Lehigh University, 5 E. Packer Ave., Bethlehem,
PA 18015 \\
$^3$Ioffe Institute, 26 Polytekhnicheskaya, St. Petersburg, 194021,
Russia \\}

\date{\today}

\begin{abstract}
Carbon nanotubes (CNTs) have large intrinsic carrier mobility due to
weak acoustic phonon scattering. However, unlike two-dimensional
metal-oxide-semiconductor field effect transistors (MOSFETs),
substrate surface polar phonon (SPP) scattering has a dramatic
effect on the CNTFET mobility, due to the reduced vertical
dimensions of the latter. We find that for the Van der Waals
distance between CNT and an SiO$_2$ substrate, the low-field
mobility at room temperature is reduced by almost an order of
magnitude depending on the tube diameter. We predict additional
experimental signatures of the SPP mechanism in dependence of the
mobility on density, temperature, tube diameter, and CNT - substrate
separation.
\end{abstract}
 \maketitle

Semiconducting carbon nanotubes (CNTs) show promise for
technological applications in electronics and optoelectronics
primarily due to weak acoustic phonon carrier scattering and a
direct bandgap \cite{Avouris}. The low-field mobilities in CNTs have
been measured by several groups \cite{Fuhrer,McEuen,Kim}. At low
biases only acoustic phonons contribute to the scattering in CNTs
and the scattering is weak \cite{Ando,Pennington, Perebeinos1}. The
CNT diameter $d$ scaling of the low-field mobility $\mu$ was
predicted\cite{Perebeinos1} to be $\mu\propto d^2$  and confirmed
experimentally \cite{McEuen}. However, the magnitude of the
experimental mobility was found to be an order of magnitude smaller
than the theoretical prediction. Recently it was shown that surface
polar phonon (SPP) scattering can be strong in CNTs lying on polar
substrates such as SiO$_2$ \cite{Rotkin}. While the SPP scattering
is of lesser importance in the conventional two-dimensional MOSFETs
\cite{Fischetti}, as we show below it is much more prominent in CNTs
due to the much smaller vertical dimension of the devices given by
the CNT diameter.

\begin{figure}[h!]
\includegraphics[height=2.35in]{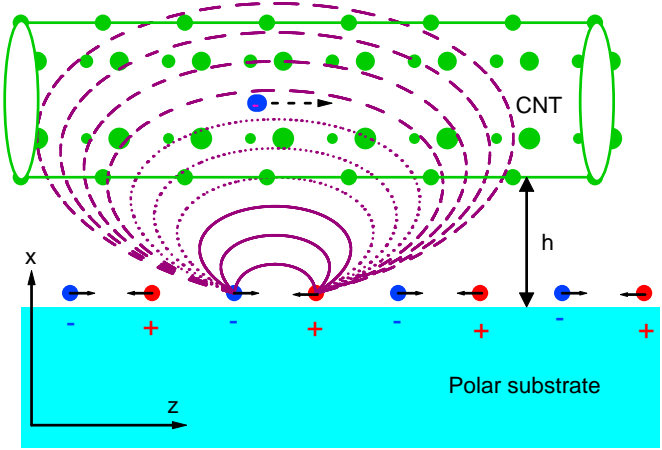}
\caption{\label{toc} Schematics of the SPP scattering in CNTs.}
\end{figure}

In this Letter we evaluate the transport properties of CNTs due to
the extrinsic scattering mechanism by surface phonons in the
insulating polar substrates as schematically shown in Fig.~\ref{toc}.
The surface phonons on polar substrates
produce a surface electric field coupled to the electrons on the
nearby CNTs. We find that SPP scattering lowers dramatically the
low-field mobility, even at room temperature. What is more
surprising, the SSP scattering can increase the drift velocity under
high bias conditions, especially at high doping levels and elevated
CNT phonon temperatures.

The interaction potential is given by \cite{Wang}:
\begin{eqnarray}
V_{int}&=&e\frac{\vec{R}-\vec{r}}{\left|\vec{R}-\vec{r}\right|^3}\vec{P}(\vec{R})
\label{eqf2}
\end{eqnarray}
where $\vec{R}$ and $\vec{r}$ are substrate phonon and CNT electron
coordinates, respectively. The strength of the SPP scattering is
given by the polarization field  $\vec{P}(\vec{R})$ at the surface
associated with a particular excited phonon mode:
\begin{eqnarray}
\vec{P}(\vec{R})&=&\sum_{\vec{q},\nu}F_{\nu}\sqrt{\frac{|\vec{q}|}{A}}
e^{\left(|\vec{q}|x+i\vec{q}\vec{R}\right)}
\left({\hat{q}}-i{\hat{x}}\right)a^{\dagger}_{-\vec{q},\nu}+h.c.
\label{eqf2a}
\end{eqnarray}
where $a^{\dagger}_{\vec{q},\nu}$ is a creation operator of a
surface phonon mode $\nu$ with two-dimensional wavevector
$\vec{q}$. The vector normal to the solid (at $x< 0$) is $\hat{x}$
and the unit vector along $\vec{q}$ is $\hat{q}$. The
normalization surface area $A$ drops out from the final result.
The magnitude of the polarization field is controlled by the
Fr${\rm \ddot{o}}$hlich coupling $F_{\nu}$:
\begin{eqnarray}
F^2_{\nu}=\frac{\hbar\omega_{SO,\nu}}{2\pi}\left(\frac{1}{\epsilon_{\infty}+1}-\frac{1}{\epsilon_{0}+1}\right),
\label{eqf1}
\end{eqnarray}
where $\hbar\omega_{SO,\nu}$ is a surface phonon energy and
$\epsilon_0$ and $\epsilon_{\infty}$ are the low- and
high-frequency dielectric constants of the polar substrate
\cite{note1}. In the following we evaluate scattering in CNTs on
SiO$_2$ surfaces, which have five phonon modes
$\hbar\omega_{SO,\nu=1...5}=50, 62, 100, 149, 149$ meV, with
Fr${\rm \ddot{o}}$hlich couplings $F^2_{\nu=1...5}=0.042, 0.38,
0.069, 1.08, 1.08$ meV respectively \cite{paramsio2}.

The CNT electron-polar surface phonon matrix element has been
derived in the envelope wavefuncton approximation in \cite{Rotkin}:
\begin{eqnarray}
\left\vert M_{kq}^{\nu}\right\vert^2&=&
\left\vert\left\langle\Psi_k\vert V_{int} \vert\Psi_{k+q}\right\rangle\right\vert^2
\nonumber \\
&\approx&\frac{4\pi e^2F^2_{\nu}}{L}I^2_{\Delta
m}\left(\frac{q_zd}{2}\right)K_{2\Delta m}\left((d+2h)q_z\right)
\label{eqf3}
\end{eqnarray}
where $\Psi_k$ and $\Psi_{k+q}$ are electron wavefunctions in CNT
with momenta $k$ and $k+q$ charaterized by the quantized angular
momentum $m$ and a continuous $z$ component along the tube axis, $d$
is a tube diameter, and $h$ is the height from the CNT surface to
the insulating substrate, see Fig.~\ref{toc}. The length of the tube
$L$ determines the k-point sampling along the tube axis $\Delta
q_{z}=2\pi/L$. The functions $I$ and $K$ are the Bessel functions of
the first and second order, respectively. It was shown that SPP
scattering is strongest when the azimuthal angular momentum is
conserved, $\Delta m=0$ \cite{Rotkin}. Therefore, to analyze the
results of the calculations it is instructive to consider the
asymptotic form of the scattering potential for $\Delta m=0$ and
large $q_z$:
\begin{eqnarray}
\left\vert M_{kq}^{\nu}\right\vert^2&\approx& \frac{4\pi
e^2F^2_{\nu}}{L}\frac{e^{-2hq_z}}{q_zd\sqrt{2\pi(d+2h)}}
\label{eqf3a}
\end{eqnarray}
This form of the SPP coupling also works well for the intermediate
values of $q_z$.

 The electron-CNT phonon scattering
is described using the Su-Schriefer-Heeger model as in ref.
\cite{Perebeinos1} with the coupling constant $g=5.3$ eV/\AA. The
electron band structure is described by the $\pi$-orbital
tight-binding model with $t_0=3$ eV.

We study the effect of the SPP scattering on transport in the
diffusive regime. We calculate the mobility by solving the
steady-state multiband Boltzmann transport equation (BTE) in the
presence of an electric field. The temperature dependent scattering
rates $W_{k,k+q}$ between electronic states $k$ and $k+q$ entering
in the BTE equation are evaluated according to:
\begin{eqnarray}
W_{k,k+q}=\frac{2\pi}{\hbar}\sum_{\nu}\left\vert M_{kq}^{\nu}\right\vert^2[
n_{q\nu}\delta\left(\varepsilon_k-\varepsilon_{k+q}+\hbar\omega_{q\nu}\right)
\nonumber \\
+\left(1+n_{-q\nu}\right)\delta\left(\varepsilon_k-\varepsilon_{k+q}-\hbar\omega_{-q\nu}\right)]
\label{eqf4}
\end{eqnarray}
where the phonon occupation number $n_{q\nu}$ is given by the
Bose-Einstein distribution. The phonon index $\nu$ runs over all
phonon modes involved in scattering, i.e. six CNT phonon modes and
five surface phonon modes in SiO$_2$, the latter are treated here as
being dispersiveless.

\begin{figure}[h!]
\includegraphics[height=2.35in]{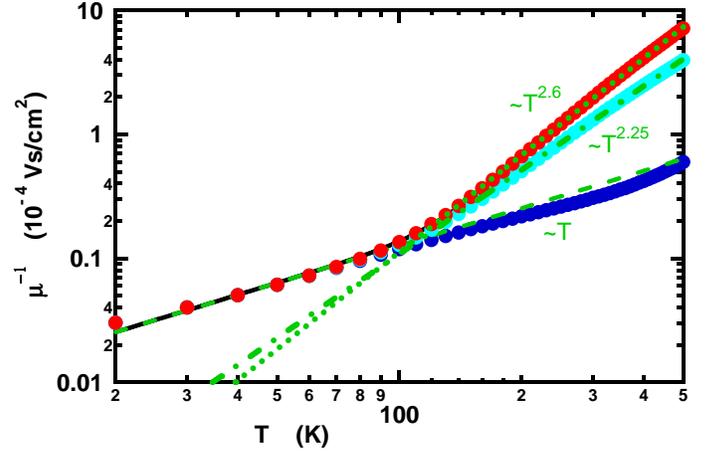}
\caption{\label{Fig1} Temperature dependence of the low-field, low density inverse mobility in
(19,0) tube ($d=1.5$ nm) on a log-log scale. The red circles are in the presence of both CNT phonons and the
SPP scattering, the cyan circles with CNT phonons and one surface phonon at $\hbar\omega_2=62$ meV,
blue circles with the CNT phonon scattering alone. The
black solid curve is a fit to Eq.~(\protect{\ref{eq1}}) with  $\alpha_{CNT}=0.0013$ and $\alpha_{SPP}=13.4$.  The low
temperature mobility is influenced by the CNT acoustic phonon scattering and gives linear
temperature dependence shown by the green dashed line. The high temperature inverse mobility can be approximated
by the power-low
 fits shown by the green dashed ($T^{2.6}$) and dashed-dotted lines ($T^{2.25}$).}
\end{figure}

The inverse mobility at low-field and low density is shown in
Fig.~\ref{Fig1} as a function of temperature. At low temperatures,
the inverse mobility is linear in temperature, because only CNT
acoustic phonon scattering dominates transport in that regime. At
temperatures above about T=100 K, SPP scattering is activated and
starts to dominate transport. The temperature dependence of the
low-field mobility becomes superlinear, due to the activated nature
of the SPP scattering and can well be approximated by the following
formula:
\begin{eqnarray}
\mu_0^{-1}=\alpha_{CNT}T+\alpha_{pol}\left(n_B(\hbar\omega_2)+\beta n_B(\hbar\omega_{4,5})\right),
\label{eq1}
\end{eqnarray}
where $\alpha_{CNT}$ is a contribution from the acoustic CNT phonons
and $\alpha_{SPP}$ is proportional to the SPP scattering rate, $n_B$
is the occupation number of SPP with energies $\hbar\omega_2=62$
meV, $\hbar\omega_{4,5}=149$ meV, and
$\beta=(F_{4}^2+F_{5}^2)/F^2_{2}=5.6$ is the ratio of the
electron-polar phonon coupling strengths fixed by the ratio of the
Fr${\rm \ddot{o}}$hlich couplings. Since the occupation number of
the highest energy SPP phonon of 149 meV is much smaller than the
$\hbar\omega_{2}=$62 meV mode, in a zeroth order approximation the
mobility below room temperature can be calculated through a coupling
only to the 62 meV SPP phonon mode. We note however that this
approximation is not exact and at least the two lowest energy modes
$\hbar\omega_{1}$ and $\hbar\omega_{2}$  have to be included to get
an adequate mobility at room temperatures. The single mode
$\hbar\omega_{2}$ calculations shown in Fig.~\ref{Fig1} (cyan color)
overestimate mobility by 50\% at room temperature.

Despite the fact that the two low energy SPP phonons at room
temperature have small occupations of $n_B=0.15$ and $0.09$
respectively, they do dominate the mobility, because of  much
stronger coupling than that to CNT acoustic phonons. The mobility at
room temperature degrades by an order of magnitude in the presence
of the SPP mechanism, depending on the tube diameter and its
distance from the substrate surface $h$. The latter depends
ultimately on the roughness of the surface. The low-field mobility
dependence on $h$ can be rationalized by  Matthiessen's rule, with
the SPP scattering rate given by Eq.~(\ref{eqf3a}) and CNT phonon
scattering described by the mobility $\mu_{CNT}$ in isolated
(suspended) CNT:
\begin{eqnarray}
\mu=\frac{\mu_{CNT}}{1+\left(\mu_{CNT}/\mu_{SiO_2}-1\right)e^{(-2q_z (h-h_0))}
\sqrt{\frac{d+2h_0}{d+2h}}},
\label{eq2}
\end{eqnarray}
where $\mu_{SiO_2}$ is mobility on polar surface substrates with the
minimum distance determined by the Van der Waals interaction
$h_0=0.35$ nm. The electron momentum transfer along the tube axis in
the scattering of a surface phonon is given by $\hbar
v_Fq_z=\sqrt{\hbar\omega_2(2\Delta+\hbar\omega_2)}$, where
$2\Delta\approx0.9/d$~eV  nm is a bandgap and $v_F\approx 10^8$ cm/s
is the Fermi velocity. We find that the ratio of the low field
mobilities at room temperature can be as large as a factor of 8.6 in
2.5 nm diameter tubes.

\begin{figure}[h!]
\includegraphics[height=2.35in]{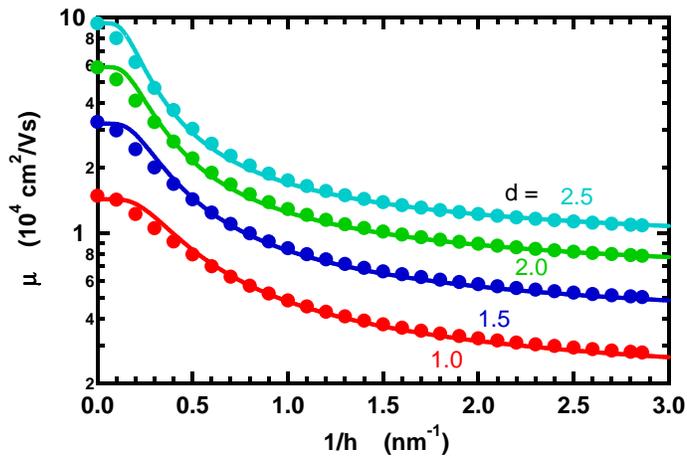}
\caption{\label{Fig2} Dependence of the low-field, low-density mobility
on the inverse distance from the polar
substrate at room temperature on a logarithmic scale.
The mobility is reduced by a factor of $\mu_{CNT}/\mu_{SiO_2}=$ 5.3, 6.5, 7.5, and 8.6 correspondingly
in (13, 0) - red, (19,0) - blue, (25, 0) - green, and (31, 0) - cyan tubes. The solid curves are calculations
according to Eq.~(\protect{\ref{eq2}}) with $q_{z}$=0.38, 0.32, 0.28, and 0.25 nm$^{-1}$.}
\end{figure}

\begin{figure}[h!]
\includegraphics[height=2.35in]{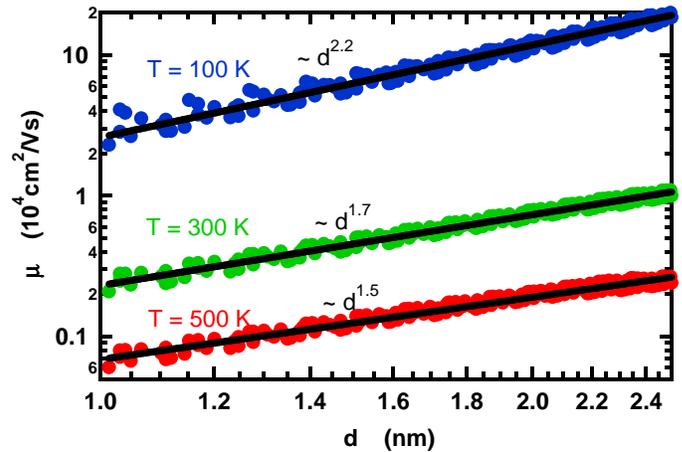}
\caption{\label{Fig3} Low-field, low-density mobility as a function of tube diameter at three temperatures
on a log-log plot for all chirality tubes. The black solid lines are best fits to
Eq.~(\protect{\ref{eq3}}) with $(A,\alpha)=$(2.6,2.18), (0.23,1.67), and (0.069,1.47) for
T = 100 K - blue, T = 300 - green, and T = 500 - red respectively.}
\end{figure}

The diameter dependence of the low-field, low-density mobility shows
a power law behavior:
\begin{eqnarray}
\mu (d)=A d^{\alpha},
\label{eq3}
\end{eqnarray}
with power $\alpha$ being temperature dependent as shown in
Fig.~\ref{Fig3}. At low temperatures, only acoustic phonons in the
CNT contribute to the scattering and the power  is $\alpha \approx
2$ as found in ref. \cite{McEuen,Perebeinos1} and Fig.~3 in
Supporting Information. At higher temperatures, SPP scattering
contributes to mobility with a weaker diameter dependence than that
for CNT phonons. This results in smaller $\alpha$ in Eq.~(\ref{eq3})
at elevated temperatures. Therefore, the relative contribution to
the total mobility of the SPP phonons in Fig.~\ref{Fig2} is larger
in larger diameter tubes, i.e $\mu_{CNT}/\mu_{SiO_2}\propto d^{0.5}$
at T=300 K and $\mu_{CNT}/\mu_{SiO_2}\propto d^{0.7}$ at T=500 K.

\begin{figure}[h!]
\includegraphics[height=3.7in]{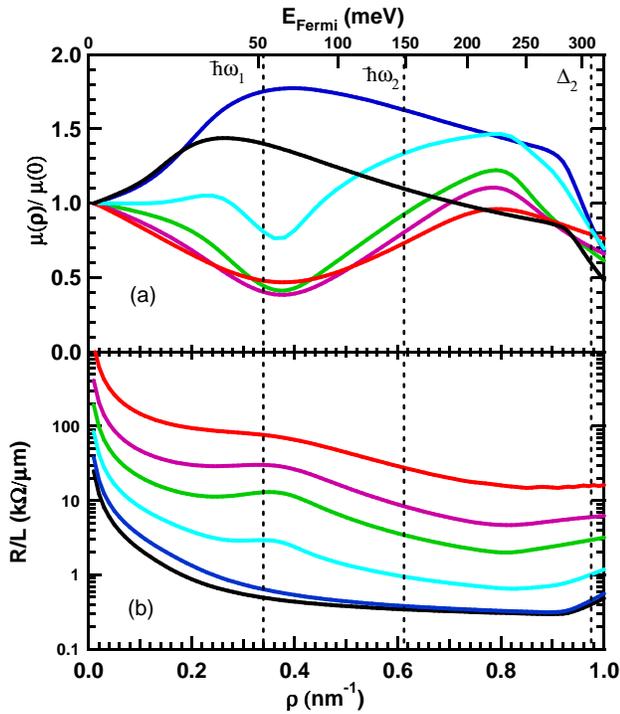}
\caption{\label{Fig4} (a) Carrier density dependence of the low-field mobility in the same tube normalized to its value at
zero density. The
vertical lines shows energies of the two polar optical phonons and
the second subband, at which scattering time is being reduced.
The upper axis shows the energy of the Fermi levels for a given density at
zero temperature.(b) Resistance per unit length in (19,0) tube
($d=1.5$ nm) at different temperatures 30 K (black), 50 K (blue),
100 K (cyan), 150 K (green), 200 K (magenta), 300 K (red) on a log scale.}
\end{figure}

Although it is natural to analyze low-density mobility
theoretically, it is difficult to measure it experimentally because
the device is close to the off-state with a high resistance. In the
case of only CNT phonon scattering, as the device is
electrostatically or chemically doped with carriers, the mobility
was shown to increase due to the smaller density of states available
for scattering as the Fermi level moves away from the bottom of the
conduction or valance bands \cite{Perebeinos2}. As the Fermi level
approaches the second band, a new channel for scattering opens up
and the mobility drops showing a non-monotonic behavior. In the
presence of the SPP scattering, the density dependence of the
mobility shows an additional minimum, when the Fermi level
approaches the lowest energy SPP phonon energy $\hbar\omega_2$ with
the strongest coupling, as shown in Fig.~\ref{Fig4}a. The size of
the mobility modulation depends on the temperature $T$, which
determines the Pauli blocking factors $(1-g_k)\approx
e^{-E_{Fermi}/k_BT}$, where $g_k$ is the distribution function from
the BTE solution and $E_{Fermi}$ is the quasi-Fermi level. At low
temperatures $T\le 100$ K, the Pauli blocking principle does not
allow any SPP scattering even if the energy conservation law can be
satisfied $E_{Fermi}\ge \hbar\omega_1$. At higher temperatures the
mobility minimum broadens out with increasing temperature. As the
density further increases, the mobility increases until it reaches
the bottom of the second band, showing a second minimum associated
with the higher subband scattering \cite{Perebeinos2}.

We have calculated the dependence of the resistivity on carrier
density for the same tube and the results are shown in
Fig.~\ref{Fig4}b. The resistivity (resistance per CNT tube length)
is non-monotonic with a minimum corresponding to occupation of the
second subband at low temperatures and a second minimum emerging at
intermediate temperatures $T\approx 100$ K due to the SPP
scattering. The prediction of the appearance of a second minimum in
the resistivity at higher temperatures could provide an experimental
signature of the SPP mechanism. On the other hand, a deviation of
the temperature dependence of the low-field mobility from the $1/T$
law is a more subtle evidence for the SPP mechanism, since the
position of the threshold voltage can be temperature dependent due
to the hysteresis \cite{Temphyster1,Temphyster2}. In addition, the
maximum mobility associated with the onset of the scattering in the
second subband shows a superlinear temperature dependence
($\mu\propto T^{1.5}$) even for only CNT phonon scattering see
Fig.~1 in Supporting Information. At the same time in metallic
tubes, the density of states around the neutrality point is constant
and, therefore, a much weaker density dependence of the mobility (or
conductance) is expected, which makes the temperature dependence of
the mobility to be a more suitable test for the SPP mechanism.
Indeed, in metallic tubes, a non-linear temperature dependence of
the resistivity has been reported \cite{Kim}.

\begin{figure}[h!]
\includegraphics[height=3.70in]{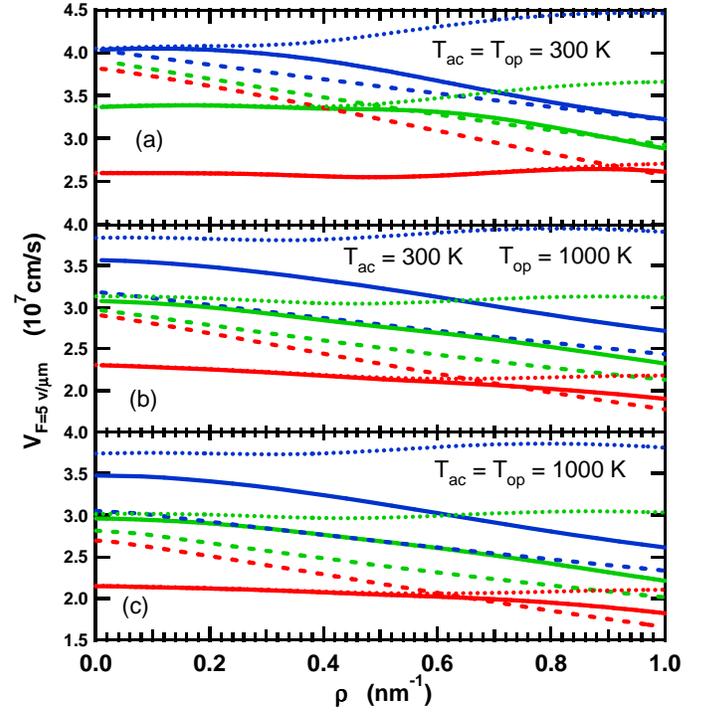}
\caption{\label{Fig5} High field velocity (at F=5 V/$\mu$m) as a function of carrier density
in three tubes (13, 0) d=1.0 nm - red, (19, 0) d=1.5 nm - green, (25, 0) d=2.0 nm - blue.
Solid curves are obtained with both CNT and SiO$_2$ phonon scattering, dashed curves with
only CNT phonon scattering, and dotted curves with both CNT and SiO$_2$ phonon scattering, but
carries restricted to the first subband. Polar phonons are kept at T = 300 K, while CNT acoustic and optical
phonons are allowed to heat up to (a) T$_{ac}$ = 300 K T$_{op}$ = 1000 K, (b) T= 1000 K and
(c) T$_{ac}$ = T$_{op}$ = 1000 K respectively. }
\end{figure}

Finally, we discuss the transport properties of semiconducting CNTs
under the high bias conditions. At low-densities with only CNT
phonon scattering, the drift velocity $v_{high}$ was found to
saturate with the field $F$ at about half the Fermi velocity
$v_{sat}\approx 5\times$ 10$^7$ cm/s (for the choice of $t_0$=3 eV):
$v_{high}^{-1}=(\mu_0 F)^{-1}+v_{sat}^{-1}$ \cite{Perebeinos1}. As
the density increases, the high field velocity drops and  density
dependence of the velocity does not show a simple saturated form
\cite{Perebeinos2}. Here we report high field velocity calculated at
a constant $F=5$ V/$\mu$m, which can be achieved in the experiments
before the long channel CNT burns under high bias condition
\cite{Chen,Eric,Pop}. The results are shown in Fig.~\ref{Fig5}. The
reference dashed lines are calculations with only CNT phonon
scattering. As the SPP scattering process is turned on, the high
bias velocity drops at small densities, especially in smaller
diameter tubes, while it increases at larger densities and larger
diameter tubes as shown in Fig.~\ref{Fig5}a. This is due to both the
hyperbolic dispersion of the CNT bandstructure \cite{Mintmire} and
the multiband nature of transport at high biases. The SPP scattering
is much stronger than the CNT optical phonon scattering and,
therefore, the tail of distribution function or effective electronic
temperature \cite{Perebeinos3} ($T_e\sim e F \lambda_{ph}$) is
expected to be smaller in the presence of the SPP mechanism, due to
the shorter mean free path $\lambda_{ph}$. We find that at low
densities, first subband dominates transport by using restricted
calculations with electrons occupying only the first subband as
shown in Fig.~\ref{Fig5} (dotted curves). As the density increases,
the band velocity increases and so does the drift velocity. In the
multiband calculations, the higher subbands with higher effective
masses and lower band velocities influence transport, which lowers
the drift velocity in Fig.~\ref{Fig5} (solid curves). In the
presence of only CNT phonon scattering, even at low densities the
high bias transport is influenced by higher subbands, which makes
the quantitative interpretation to be more difficult although it
shows similar trends.

It has been suggested that under high bias conditions the optical
phonons can be out of equilibrium with the acoustic CNT phonons
\cite{Lazzeri,Pop}. Recently, the non-equilibrium phonon population
in CNTs under high bias conditions was directly observed
experimentally \cite{Steiner}. To study the influence of the hot
phonon effect on the transport properties, we calculate the drift
velocity as a function of density for three different tubes in
Fig.~\ref{Fig5}b. In the case of only CNT scattering, we find a
significant drop in the drift velocity with the elevation of the
optical phonon temperature. This drop gives rise to the negative
differential conductance as observed in the suspended CNTs
\cite{Pop}. On the other hand, the magnitude of the drift velocity
change in the presence of the SPP is much smaller, because SPP,
which dominates transport, are kept at ambient temperature. The
weaker effect of the CNT phonons on transport properties leads to
near disappearance of the negative differential conductance even
when CNT optical phonon temperature is elevated \cite{Steiner}.
Finally, we find that the drift velocity under high bias conditions
in the presence of the SPP scattering is not effected by the
acoustic CNT phonon temperature as is seen from Fig.~\ref{Fig5}c,
which is essentially identical to Fig.~\ref{Fig5}b.

In conclusion, we calculated the effect of the surface polar phonon
scattering on transport in semiconducting CNTs. We find that at room
temperature SPP scattering reduces the mobility by an order of
magnitude from the intrinsic value determined by the CNT acoustic
phonon scattering. The diameter dependence of the mobility in the
presence of SPP can still be well approximated by a power low, but
with a reduced temperature dependent value of the power as the
relative role of the SPP scattering on transport increases with
increasing tube diameter. At temperatures above about 100 K, we
predict an additional minimum in the mobility-density dependence
associated with the low energy SPP phonon scattering. This can serve
as an experimental signature of the SPP mechanism. Under high bias
conditions we find, counter-intuitively, that SPP scattering  may
increase the drift velocity, especially in large diameter tubes,
high carrier density, and elevated CNT optical phonon temperatures
due to the non-parabolicity of the bandstructure and influence of
the higher subbands.

Finally, it is useful to comment on the SPP scattering in CNTs on
other possible polar gate insulators, such as Al$_2$O$_3$, AlN,
ZrO$_2$, HfO$_2$, and ZrSiO$_4$ in which the lowest energy phonon
with an appreciable coupling is estimated to be
$\hbar\omega_{SO}=\hbar\omega_{TO}\sqrt{(1+\epsilon_0)/(1+\epsilon_{\infty})}=$
87, 108, 37, 24, 60 meV, respectively using the parameters in ref.
\cite{Fischetti}. The Fr${\rm \ddot{o}}$hlich couplings
Eq.~(\ref{eqf1}) to the low energy modes in these materials are
similar to that in SiO$_2$. Therefore, we may expect that from the
point of SPP scattering, AlN would be the best choice for the
insulator, while HfO$_2$ and ZrO$_2$ would give the strongest
scattering at room temperature.

\newpage


\begin{thebibliography}{}
\bibitem{note1} We addopt a back-gate geometry, with one-dimensional
CNT lying on two-dimensional polar substrate. Such that we
approximate a 2D top dielectric, i.e. CNT in air, by $\epsilon=1$.
\end{thebibliography}
\end{document}